\documentclass[smallextended,envcountsame]{svjour3}

\usepackage{todonotes}

\usepackage{amsmath}
\usepackage{tikz}
\usepackage{amssymb}
\usepackage{subfig}
\usepackage{tcolorbox}
\usepackage{listings}

\def\dosth#1{\ifx###1##\else\dofirst#1\anytoken\fi}
\def\doagain#1\anytoken{\dosth{#1}}
\def\payoffpairs#1#2#3{\m=#1\multiply\m by 4 \advance\m by -1 \n=1
  \def\dofirst##1{\put(\n,-\m){\makebox(0,0){\strut##1}}\advance\n by 4 \doagain}%
  \dosth{#2\strut}%
  \m=#1\multiply\m by 4 \advance\m by -3 \n=3 \dosth{#3\strut}}
\def\singlepayoffs#1#2{\m=#1\multiply\m by 4 \advance\m by -2 \n=2
  \def\dofirst##1{\put(\n,-\m){\makebox(0,0){\strut##1}}\advance\n by 4 \doagain}%
  {\large\dosth{#2\strut}}}
\newcommand{\bimatrixgame}[8]{%
\setlength{\unitlength}{#1}%
\newcount\rows
\newcount\cols
\rows=#2
\cols=#3
\newcount\rowcoord
\newcount\colcoord
\rowcoord=\rows
\colcoord=\cols
\multiply\rowcoord by 4
\multiply\colcoord by 4
\newcount\m
\newcount\n
\m=\rowcoord
\n=\colcoord
\advance\m by 2 % 2 units left of payoff table
\advance\n by 2 % 2 units above payoff table
\begin{picture}(\n,\m)(-2,-\rowcoord)
\m=\rows
\n=\cols
\advance\m by 1
\advance\n by 1 
\thinlines
\multiput(0,0)(0,-4){\m}{\line(1,0){\colcoord}}
\multiput(0,0)(4,0){\n}{\line(0,-1){\rowcoord}}
\put(0,0){\line(-1,1){2}}
\put(-1.5,0.5){\makebox(0,0)[r]{#4}}  % name player I
\put(-.7,1.7){\makebox(0,0)[l]{#5}}   % name player II
\n=2
\def\dofirst##1{\put(-0.8,-\n){\makebox(0,0)[r]{\strut##1}}\advance\n by 4
   \doagain}
\dosth{#6\strut} 
\n=2
\def\dofirst##1{\put(\n,1.0){\makebox(0,0){\strut##1}}\advance\n by 4
   \doagain}
\dosth{#7\strut}#8%
\end{picture}}
\newcommand{\rats}{\mathbb{R}}

\newcommand{\x}{\mathbf{x}}
\newcommand{\y}{\mathbf{y}}
\newcommand{\yimp}{\mathbf{y}^{\text{imp}}}

\newcommand{\row}{i}
\newcommand{\rowx}{{i'}}
\newcommand{\rowb}{{\bar{\imath}}}
\newcommand{\col}{j}
\newcommand{\colx}{{j'}}
\newcommand{\qb}{\bar{q}}

\newcommand{\dbb}{\mathbf{bb}}
\newcommand{\dbs}{\mathbf{bs}}
\newcommand{\dbo}{\mathbf{bo}}
\newcommand{\dsb}{\mathbf{sb}}
\newcommand{\dss}{\mathbf{ss}}
\newcommand{\dso}{\mathbf{so}}
\newcommand{\dob}{\mathbf{ob}}
\newcommand{\dos}{\mathbf{os}}
\newcommand{\doo}{\mathbf{oo}}
\newcommand{\dq}{\mathbf{q}}
\newcommand{\dqb}{\mathbf{\bar{q}}}

\newcommand{\third}{$\frac{1}{3}$}

\newcommand{\lp}{LP}
\newcommand{\bestrow}{BestR}
\newcommand{\bestcol}{BestC}

\DeclareMathOperator{\prob}{Pr}
\DeclareMathOperator{\supp}{Supp}
\DeclareMathOperator{\sol}{Sol}

\begin{document}

\title{Approximate Well-supported Nash Equilibria Below
Two-thirds\thanks{This work was supported by by EPSRC grants
EP/H046623/1, EP/G069239/1, EP/G069034/1, and EP/L011018/1.}}

\author{John Fearnley
\and
Paul W. Goldberg
%\thanks{Supported by EPSRC grant EP/G069239/1 ``Efficient Decentralised Approaches in Algorithmic Game Theory.''} 
\and
Rahul Savani
\and
Troels Bjerre S\o rensen}
%\thanks{Supported by EPSRC grant ``Efficient Decentralised Approaches in Algorithmic Game Theory''}

\institute{J. Fearnley \and R. Savani \at Department of Computer Science, University of Liverpool, UK. 
\and
P. Goldberg \at Wolfson Building, Parks Road, University of Oxford, UK. 
\and
T.B. S\o rensen \at ITU Copenhagen, Rued Langgaards Vej 7, 2300 K\o benhavn S,
Denmark.}

\maketitle

\begin{abstract}
In an $\epsilon$-Nash equilibrium, a player can gain at most $\epsilon$ by
changing his behaviour. Recent work has addressed the
question of how best to compute $\epsilon$-Nash equilibria, and for what
values of $\epsilon$ a polynomial-time algorithm exists.
An $\epsilon$-{\em well-supported} Nash equilibrium ($\epsilon$-WSNE) has the
additional requirement that any strategy
that is used with non-zero probability by a player must have payoff at most
$\epsilon$ less than a best response. 
A recent algorithm of Kontogiannis and Spirakis shows how to compute a
$2/3$-WSNE in polynomial time, for bimatrix games.
Here we introduce a new technique that leads to
an improvement to
the worst-case approximation guarantee.
\end{abstract}

%\begin{keyword}
%Bimatrix games, Nash equilibria, well-supported approximate equilibria.
%\end{keyword}

\section{Introduction}

In a bimatrix game, a Nash equilibrium is a pair of strategies in which the two
players only assign probability to best response strategies. The
apparent hardness of computing an exact Nash equilibrium~\cite{DGP,CDT} has led
to work on computing approximate Nash equilibria, and two notions of approximate
Nash equilibria have been developed. The first, and more widely studied, notion
is of an \emph{$\epsilon$-approximate Nash equilibrium} ($\epsilon$-Nash), where
each player is required to achieve an expected payoff that is within~$\epsilon$
of a best response. A line of work~\cite{DMP,Progress,Bosse} has investigated
the value of $\epsilon$ that can be guaranteed in polynomial time. The current
best result in this setting is a polynomial time algorithm that always finds a
0.3393-Nash equilibrium~\cite{TS}.

However,~$\epsilon$-Nash equilibria have a drawback: since they only require
that the expected payoff is within~$\epsilon$ of a pure best response, it is
possible that a player could be required to place probability
on a strategy that is arbitrarily far from being a best response. This issue is
addressed by the second notion of an approximate Nash equilibrium. An
\emph{$\epsilon$-well supported approximate Nash equilibrium} ($\epsilon$-WSNE),
requires that both players only place probability on strategies that have payoff
within $\epsilon$ of a pure best response. This is a stronger notion of
equilibrium, because every $\epsilon$-WSNE is an $\epsilon$-Nash, but the
converse is not true.

In contrast to $\epsilon$-Nash, there has been relatively little work
$\epsilon$-WSNE. The first result on the subject gave a $\frac{5}{6}$ additive
approximation~\cite{DMP}, but this only holds if a certain a graph-theoretic
conjecture is true. The best-known polynomial-time additive approximation
algorithm was given by Kontogiannis and Spirakis, and achieves a
$\frac{2}{3}$-approximation~\cite{KS}. We will call this algorithm the KS
algorithm. In~\cite{KS07}, which is an earlier conference version of~\cite{KS},
the authors presented an algorithm that they claimed was polynomial-time and
achieves a $\phi$-WSNE, where $\phi = \frac{\sqrt{11}}{2} - 1 \approx 0.6583$,
but this was later withdrawn, and instead the polynomial-time
$\frac{2}{3}$-approximation algorithm was presented in~\cite{KS}. 
Recently, it has been shown that for every $\delta > 0$, a
$(\frac{1}{2}+\delta)$-WSNE can be found in polynomial times for
\emph{symmetric} bimatrix games~\cite{CFJ14}. It has also been shown that there
is a PTAS for $\epsilon$-WSNE if and only if there is a PTAS for
$\epsilon$-Nash~\cite{CDT}.

%\begin{figure}
%\hspace{0.1\textwidth}
%\subfloat[]
%%[A bimatrix game for which the KS algorithm produces a $\frac{2}{3}$-WSNE.]
%{
	%\label{fig:ex1}
	%\bimatrixgame{3.7mm}{2}{2}{I}{II}%
	%{{$T$}{$B$}}%
	%{{$\ell$}{$r$}}%
	%{
	%\payoffpairs{1}{{\third$-\tau$}{$1$}}{{$1$}{\third$-\tau$}}
	%\payoffpairs{2}{{$\tau$}{$\tau$}}{{$0$}{$0$}}
	%}
%}
%\hfill
%\subfloat[]
%%[A game in which the solution provided by the KS algorithm cannot be improved through rearranging probabilities.]
%{
	%\label{fig:ex2}
	%\bimatrixgame{3.7mm}{3}{2}{I}{II}%
	%{{$T$}{$M$}{$B$}}%
	%{{$\ell$}{$r$}}%
	%{
	%\payoffpairs{1}{{\third$-\tau$}{$1$}}{{$1$}{\third$-\tau$}}
	%\payoffpairs{2}{{$1$}{\third$-\tau$}}{{\third$-\tau$}{$1$}}
	%\payoffpairs{3}{{$\tau$}{$\tau$}}{{$0$}{$0$}}
	%}
%}
%\hspace{0.1\textwidth}
%\caption{Two examples that approach the worst case for the KS algorithm}
%\label{fig:ex1whole}
%\end{figure}

%In this section we give some intuition about the approach, and how we build
%on the KS algorithm.

\paragraph{\bf Our contribution} 

In this paper, we develop an algorithm for finding an $\epsilon$-WSNE with
$\epsilon < \frac{2}{3}$. Our approach to modifying the KS algorithm for finding
a $\frac{2}{3}$-WSNE, by adding two additional procedures: we perform a
brute-force search that finds the best WSNE with a $2 \times 2$ support, and we
attempt to improve the $\epsilon$-WSNE returned by the KS algorithm by shifting
the probabilities of the two players. We show that one of these two approaches
will always find an $\epsilon$-WSNE with $\epsilon = \frac{2}{3} - 0.005913759$.
Our results are particularly interesting when compared to a recent support size
lower bound of Anbalagan, Norin, Savani, and Vetta, who showed that there exist
games in which all $\epsilon$-WSNE with $\epsilon < \frac{2}{3}$ have
super-constant sized supports~\cite{ANSV13}.

A preliminary version of this paper was published in the proceedings of SAGT
2012~\cite{SAGT}. In that version of the paper, we gave a polynomial time algorithm for
finding an $\epsilon$-WSNE with $\epsilon = \frac{2}{3} - 0.004735$. It turns
out that one of the inequalities used to show this result\footnote{The
inequality in question appeared in Proposition 16 of the preliminary version,
and is now part of Proposition~\ref{prop:imptoy}.} was not as strong as it could
have been, and correcting this led to the improved bound in this version of the
paper. We have also greatly simplified the computer-assisted proof that is used
at the end of the paper. The preliminary version of the paper used a rather
opaque method involving sensitivity analysis of a linear program. We have
reformulated the LP so that the relevant values can be read directly from the
solution of the LP.

The paper will proceed as follows. In Section~\ref{sec:defs} we give the basic
definitions that will be needed in this paper. In Section~\ref{sec:outline} we
give a high level overview of our algorithms, along with the intuition behind
our two modifications. In Section~\ref{sec:ouralgorithm}, we formally define our
algorithm and state our main theorem. In Section~\ref{sec:roadmap} we give a
high level overview of the proof, before then proceeding with the proof in
Sections~\ref{sec:reks} through~\ref{sec:new}.

\section{Definitions}\label{sec:defs}

A \emph{bimatrix game} is a pair $(R, C)$ of two $n\times n$ matrices: $R$ gives
payoffs for the \emph{row player}, and $C$ gives payoffs for the
\emph{column player}. We assume that all payoffs are in the range
$[0, 1]$. We use $[n]=\{1,2,\ldots n\}$ to denote the \emph{pure
strategies} for each player. To play the game, both players simultaneously
select a pure strategy: the row player selects a row $i \in [n]$, and the column
player selects a column $j \in [n]$. The row player then receives 
$R_{i, j}$, and the column player receives $C_{i, j}$.

A \emph{mixed strategy} is a probability distribution over $[n]$. We denote a
mixed strategy as a vector $\x$ of length $n$, such that $\x_i$ is the
probability that the pure strategy~$i$ is played. The \emph{support} of mixed
strategy $\x$, denoted $\supp(\x)$, is the set of pure strategies $i$ with $\x_i
> 0$. If $\x$ and $\y$ are mixed strategies for the row and column player,
respectively, then we call $(\x, \y)$ a \emph{mixed strategy profile}. 

Let $\y$ be a mixed strategy for the column player. The \emph{best responses}
against $\y$ for the row player is the set of pure strategies that maximize
the payoff against $\y$. More formally, a pure strategy $i \in [n]$ is a best
response against $\y$ if, for all pure strategies $i' \in [n]$ we have:
$\sum_{j \in \protect [n]} \y_j \cdot R_{i, j} \ge \sum_{j \in \protect [n]}
\y_j \cdot R_{i', j}$.
Column player best responses are defined analogously. A mixed strategy
profile $(\x, \y)$ is a \emph{mixed Nash equilibrium} if every pure strategy in
$\supp(\x)$ is a best response against~$\y$, and every pure strategy in
$\supp(\y)$ is a best response against~$\x$. Nash~\cite{N} showed 
that all bimatrix games have a mixed Nash equilibrium.

An \emph{approximate well-supported Nash equilibrium} weakens
the requirements of a mixed Nash equilibrium. For a mixed strategy $\y$ of the
column player, a pure strategy $i \in [n]$ is an \emph{$\epsilon$-best response}
for the row player if, for all pure strategies $i' \in [n]$ we have: $\sum_{j
\in \protect [n]} \y_j \cdot R_{i, j} \ge \sum_{j \in \protect [n]} \y_j \cdot
R_{i', j} - \epsilon$. We define $\epsilon$-best responses for the column player
analogously. A mixed strategy profile $(\x, \y)$ is an
\emph{$\epsilon$-well-supported Nash equilibrium} ($\epsilon$-WSNE) if every
pure strategy in $\supp(\x)$ is an $\epsilon$-best response against~$\y$, and
every pure strategy in $\supp(\y)$ is an $\epsilon$-best response against~$\x$.

\section{Outline}
\label{sec:outline}

Before we give a technical presentation of our algorithm, we begin by giving the
high level ideas behind our techniques. Our approach builds upon the algorithm
of Kontogiannis and Spirakis for finding a $\frac{2}{3}$-WSNE, so let us begin
by describing their algorithm. Given a bimatrix game $(R, C)$, the KS algorithm
performs two steps:
\begin{enumerate}
\item Check if there is a pure strategy profile under which both players get
payoff at least $\frac{1}{3}$. If so, that pure strategy profile is a
$\frac{2}{3}$-WSNE.
\item Construct the zero-sum game $(D, -D)$ where $D = \frac{1}{2}(R - C)$, and
let $(\x, \y)$ be a Nash-equilibrium of $(D, -D)$.
\end{enumerate}
Kontogiannis and Spirakis showed that if step 1 failed to find a pure
$\frac{2}{3}$-WSNE of $(R, C)$, then $(\x, \y)$ is a $\frac{2}{3}$-WSNE of $(R,
C)$. Our goal is to show that the WSNEs found by the KS algorithm can be
improved: either by \emph{shifting probabilities,} or by finding a
\emph{matching pennies} sub-game. We now show the motivation behind these two
procedures.

\begin{figure}
\hspace{0.1\textwidth}
\subfloat[]
%[A bimatrix game for which the KS algorithm produces a $\frac{2}{3}$-WSNE.]
{
	\label{fig:ex1a}
	\bimatrixgame{3.7mm}{2}{2}{I}{II}%
	{{$T$}{$B$}}%
	{{$\ell$}{$r$}}%
	{
	\payoffpairs{1}{{\third}{$1$}}{{$1$}{\third}}
	\payoffpairs{2}{{$0$}{$0$}}{{$0$}{$0$}}
	}
}
\hfill
\subfloat[]
%[A game in which the solution provided by the KS algorithm cannot be improved through rearranging probabilities.]
{
	%\label{fig:ex2}
	%\bimatrixgame{3.7mm}{3}{2}{I}{II}%
	%{{$T$}{$M$}{$B$}}%
	%{{$\ell$}{$r$}}%
	%{
	%\payoffpairs{1}{{\third$-\tau$}{$1$}}{{$1$}{\third$-\tau$}}
	%\payoffpairs{2}{{$1$}{\third$-\tau$}}{{\third$-\tau$}{$1$}}
	%\payoffpairs{3}{{$\tau$}{$\tau$}}{{$0$}{$0$}}
	%}
	\label{fig:ex1b}
	\bimatrixgame{3.7mm}{2}{2}{I}{II}%
	{{$T$}{$B$}}%
	{{$\ell$}{$r$}}%
	{
	\payoffpairs{1}{{-\third}{\third}}{{\third}{-\third}}
	\payoffpairs{2}{{$0$}{$0$}}{{$0$}{$0$}}
	}
}
\hspace{0.1\textwidth}
\caption{The left figure shows a worst case example for the KS algorithm. The
right figure shows the corresponding zero-sum game $(D, -D)$.}
\label{fig:ex1}
\end{figure}

\paragraph{\bf Shifting probabilities} Figure~\ref{fig:ex1a} shows an example
for which the KS algorithm actually produces a $\frac{2}{3}$-WSNE. For
simplicity of exposition, we have ignored the first part of the algorithm here:
note that in $(T, \ell)$ and $(T, r)$ both players have payoff greater than or
equal to $\frac{1}{3}$. If we replace both $\frac{1}{3}$ payoffs with
$\frac{1}{3} - \delta$, for some arbitrarily small $\delta > 0$, then this issue
is avoided, and the properties of the example do not significantly change.

Figure~\ref{fig:ex1b} shows the corresponding zero-sum game. Let $(\x, \y)$ be a
strategy profile in which the row player plays $B$, and the column player mixes
equally between $\ell$ and $r$. Observe that $(\x, \y)$ is a Nash equilibrium of
the zero-sum game, and that it is $\frac{2}{3}$-WSNE of $(R, C)$, and no better.
Therefore, this example is a worst-case example for the KS-algorithm.

Our observation is that $(\x, \y)$ can be improved by shifting probabilities.
We can improve things for the row player by transferring some of the
{\em column} player's probability from $r$ to $\ell$.
This reduces the payoff of $T$ while leaving the payoff of $B$ unchanged. Thus,
$B$ becomes an $\epsilon$-best response for $\epsilon < \frac{2}{3}$, and we
obtain a better WSNE.

\begin{figure}
\hspace{0.1\textwidth}
\subfloat[]
{
	\label{fig:ex2a}
	\bimatrixgame{3.7mm}{3}{2}{I}{II}%
	{{$T$}{$M$}{$B$}}%
	{{$\ell$}{$r$}}%
	{
	\payoffpairs{1}{{\third}{$1$}}{{$1$}{\third}}
	\payoffpairs{2}{{$1$}{\third}}{{\third}{$1$}}
	\payoffpairs{3}{{$0$}{$0$}}{{$0$}{$0$}}
	}
}
\hfill
\subfloat[]
%[A game in which the solution provided by the KS algorithm cannot be improved through rearranging probabilities.]
{
	\label{fig:ex2b}
	\bimatrixgame{3.7mm}{3}{2}{I}{II}%
	{{$T$}{$M$}{$B$}}%
	{{$\ell$}{$r$}}%
	{
	\payoffpairs{1}{{$-$\third}{\third}}{{\third}{$-$\third}}
	\payoffpairs{2}{{\third}{$-$\third}}{{$-$\third}{\third}}
	\payoffpairs{3}{{$0$}{$0$}}{{$0$}{$0$}}
	}
}
\hspace{0.1\textwidth}
\caption{The left figure shows an example for which the approach of shifting
probabilities fails. The right figure shows the corresponding zero-sum game.}
\label{fig:ex2}
\end{figure}

\paragraph{\bf Matching pennies}
Figure~\ref{fig:ex2} shows a game in which the approach of shifting
probabilities does not work. To see this, observe that the strategy profile
$(\x, \y)$ where the row player plays $B$, and the column player mixes uniformly
over $\ell$ and $r$ is a Nash equilibrium of the game shown in
Figure~\ref{fig:ex2b}. When $(\x, \y)$ is played in the original game in
Figure~\ref{fig:ex2a}, this gives a $\frac{2}{3}$-WSNE and no better. However,
in this case, the column player cannot make the row player happier by shifting
probabilities: if probability is shifted to $\ell$, then the payoff of strategy
$T$ will increase, and if probability is shifted to $r$ then the payoff of
strategy $M$ will increase.

In this case, we use a different approach. We observe that the $2 \times 2$
subgame induced by $\ell$, $r$, $T$, and $M$, is similar to a matching pennies
game. If the row player mixes uniformly over $M$ and $T$, while the column
player mixes uniformly of $\ell$ and $r$, then both players will obtain payoff
at least $0.5$, which yields a $0.5$-WSNE.

\paragraph{\bf Our approach} We will show that one of these two techniques can
always be applied. Our algorithm will first perform a brute force search over
all $2 \times 2$ sub-games in order to determine whether there is a matching
pennies sub game. If such a game is not found, then we run the KS algorithm and
attempt to shift probabilities in the resulting strategy profile. Ultimately, we
show that this algorithm always produces a $(\frac{2}{3} - 0.005913759)$-WSNE.

\section{Our Algorithm}
\label{sec:ouralgorithm}

In this section we formally describe our algorithm for finding a WSNE. We begin
by describing a method for finding the best WSNE on a given pair of supports,
and then move on to describe the three procedures that make up our algorithm.

\paragraph{\bf The best WSNE on a pair of supports}
 Let~$S_c$ and $S_r$ be
supports for the column and row player, respectively. We first define an LP,
which assumes that the row player uses a strategy with support $S_r$, and then
finds a strategy on $S_c$ that minimizes the difference between the row player's
best response payoff, and the payoff of the strategies in $S_r$.
\begin{definition} 
\label{def:rearrangey}
For each $S_r, S_c \subseteq [n]$, we define $\bestrow(S_r, S_c)$ to be the
following linear
program with variables $\epsilon \in \rats$ and $\y \in \rats^n$:
\begin{align}
\textbf{Minimize:} && \epsilon  & \notag \\
\notag \\
\label{eqn:ylp1}
\textbf{Subject to:} && R_{\rowx} \cdot \y - R_\row \cdot \y & \le
\epsilon & \row &\in S_r, \; \rowx \in [n] \\
\nonumber
&& \y_\col  & = 0 & \col &\notin S_c \\
\nonumber
&& \y_\col &\ge 0 & \col &\in [n]  \\
\nonumber
&& \sum_{\col \in [n]} \y_\col &= 1 
\end{align}
\end{definition}

\noindent
Similarly, the following LP assumes that the column player uses a
strategy with support $S_c$, and finds a strategy on $S_r$ that minimizes the
difference between the column player's best response payoff, and the payoff of
the strategies in $S_c$.

\begin{definition}
\label{def:rearrangex}
For each $S_r, S_c \subseteq [n]$, we define $\bestcol(S_r, S_c)$ to be the
following linear
program with variables $\epsilon \in \rats$ and $\x \in \rats^n$:
\begin{align}
\textbf{Minimize:} && \epsilon  & \notag \\
\notag \\
\label{eqn:xlp1}
\textbf{Subject to:} && 
C^T_{\colx} \cdot \x - C^T_\col \cdot \x &\le \epsilon
& \col &\in S_c, \; \colx \in [n] \\
\nonumber
&& \x_\row  & = 0 & \row &\notin S_r \\
\nonumber
&& \x_\row & \ge 0 & \row & \in [n] \\
\nonumber
&& \sum_{\col \in [n]} \x_\row & = 1 
\end{align}
\end{definition}

We now prove that these two LPs give a WSNE that is at least as good as the best
WSNE on the supports $S_r$ and $S_c$. Let $(\y^*, \epsilon_\y)$ be a solution of
$\bestrow(S_r, S_c)$, let $(\x^*, \epsilon_\x)$ be a solution of $\bestcol(S_r,
S_c)$, and let $\epsilon^*$ to be $\max(\epsilon_\x, \epsilon_\y)$.

\begin{proposition}
\label{prop:troelswsne}
We have: 
\begin{enumerate}
\item $(\x^*, \y^*)$ is an $\epsilon^*$-WSNE.
\item For every $\epsilon$-WSNE $(\x, \y)$ with $\supp(\x) = S_r$ and $\supp(\y)
= S_c$, we have $\epsilon^* \le \epsilon$.
\end{enumerate}
\end{proposition}
\begin{proof}
The first claim is straightforward, because Constraint~\ref{eqn:ylp1} ensures
that every strategy $\row \in \supp(\x^*) \subseteq S_r$ is an
$\epsilon_\y$-best response against $\y^*$, and every strategy $\col \in
\supp(\y^*) \subseteq S_c$ is an $\epsilon_\x$-best response against $\x^*$.
Therefore $(\x^*, \y^*)$ is a $\epsilon^*$-WSNE.

For the second claim, let $(\x, \y)$ be an $\epsilon$-WSNE on the supports $S_r$
and $S_c$. Since every row $i \in \supp(\x) = S_r$ is an $\epsilon$-best
response against $\y$, we must have that $\y$ and $\epsilon$ are feasible in 
$\bestrow(S_r, S_c)$. For the same reason, we have that $\x$ and $\epsilon$ are
feasible in $\bestcol(S_r, S_c)$. Therefore, we must have $\epsilon^* \le
\epsilon$.
%Since $\y^*$ is a solution of the LP given in Definition~\ref{def:rearrangey},
%Constraint~\eqref{eqn:ylp1} implies that
%\begin{equation*}
%R_{\rowx} \cdot \y^* - R_\row \cdot \y^* \le \epsilon_\y,
%\end{equation*}
%for every row $\row \in \supp(\x^*)$, and every row $\rowx \in \protect [n]$.
%Therefore, $\x^*$ is an $\epsilon_\y$-best response against $\y^*$.
%Similarly, since $\x^*$ is a solution of the LP given in
%Definition~\ref{def:rearrangex},  Constraint~\eqref{eqn:xlp1} implies that
%\begin{equation*}
%C^T_{\colx} \cdot \x^* - C^T_\col \cdot \x^* \le \epsilon_\x,
%\end{equation*}
%for every column $\col \in \supp(\y^*)$, and every column $\colx \in \protect
%[n]$. Therefore, $\y^*$ is an $\epsilon_\x$-best response against $\x^*$.
%Thus, we have that $(\x^*, \y^*)$ is an $\epsilon^*$-WSNE. 
\end{proof}

Proposition~\ref{prop:troelswsne} implies that $(\x^*, \y^*)$ is at least as
good as the best WSNE on the supports $S_r$ and $S_c$. Note that it is possible
that $(\x^*, \y^*)$ may actually be better than any WSNE on these
supports, because the LPs do not require that~$\x^*$ places probability on all
strategies in $S_r$, or that $\y^*$ places probability on all strategies in
$S_c$.

%More importantly, we can show that $(\x^*, \y^*)$ is at least as good, or better
%than, all well-supported Nash equilibria with support~$S_c$ and~$S_r$. 

%\begin{proposition}
%\label{prop:awesome}
%hello
%\end{proposition}
%\begin{proof}
%Let $(\x, \y)$ be an arbitrary $\epsilon$-WSNE on the supports $S_r$ and $S_c$.
%Since $\supp(\y) = S_r$, we know that $\y$ satisfies the constraints given
%by~\eqref{eqn:ylp1}. Moreover, since $\x$ is an $\epsilon$-best response to
%$\y$, we must have, for every row $\row \in \supp(\x)$, and every row $\rowx \in
%\protect [n]$:
%\begin{equation*}
%R_{\rowx} \cdot \y - R_\row \cdot \y \le \epsilon.
%\end{equation*}
%This implies that $(\y, \epsilon)$ is feasible in the LP given by
%Definition~\ref{def:rearrangey}, which implies that $\epsilon \ge \epsilon_\y$.

%Similarly, since $\supp(\x) = S_r$, we know that $\x$ satisfies the constraints
%given by~\eqref{eqn:xlp1}. 
%Moreover, since $\y$ is an $\epsilon$-best response
%to $\x$, we must have, for every column $\col \in \supp(\y)$, and every column
%$\colx \in \protect [n]$:
%\begin{equation*}
%C^T_{\colx} \cdot \x - C^T_\col \cdot \x \le \epsilon.
%\end{equation*}
%This implies that $(\x, \epsilon)$ is feasible in the LP given by
%Definition~\ref{def:rearrangex}, which implies that $\epsilon \ge \epsilon_\x$.

%Since $\epsilon \ge \epsilon_\y$ and  $\epsilon \ge \epsilon_\x$, we must have
%$\epsilon \ge \max(\epsilon_\y, \epsilon_\x) = \epsilon^*$. 
%\end{proof}

\paragraph{\bf Our Algorithm}
We now describe our algorithm for finding a WSNE in a bimatrix game. Our
algorithm for finding a WSNE consists of three distinct procedures.

\begin{itemize}
\item \textbf{Procedure 1: find the best pure WSNE.}
The KS algorithm requires a preprocessing step that eliminates all games that
have a pure $\frac{2}{3}$-WSNE, and this is a generalisation of that step.
Suppose that the row player plays row $\row$, and that the column player plays
column $\col$. Let: $\epsilon_r = \max_\rowx (R_{\rowx, \col}) - R_{\row,
\col}$, and $\epsilon_c = \max_\colx (C_{\row, \colx}) - C_{\row, \col}$. Thus
$\row$ is an $\epsilon_r$-best response against $\col$, and that $\col$ is an
$\epsilon_c$-best response against $\row$. Therefore, $(i, j)$ is a
$\max(\epsilon_r, \epsilon_c)$-WSNE. We can find the best pure WSNE by checking
all $O(n^2)$ possible pairs of pure strategies. Let $\epsilon_p$ be the best
approximation guarantee that is found by this procedure.
 
\item \textbf{Procedure 2: find the best WSNE with $2 \times 2$ support.}
We can use the linear programs from Definitions~\ref{def:rearrangey}
and~\ref{def:rearrangex} to implement this
procedure. For each of the $O(n^4)$ possible $2 \times 2$ supports, we solve the
LPs to find a WSNE. Proposition~\ref{prop:troelswsne} implies that this WSNE is at
least as good as the best WSNE on those supports.
Let $\epsilon_m$ be the best approximation guarantee that is found by this procedure.

\item \textbf{Procedure 3: find an improvement over the KS algorithm.}
The KS algorithm finds an exact Nash equilibrium $(\x, \y)$ of the zero-sum game $(D, -D)$,
where $D=\frac{1}{2}(R-C)$. 
%Kontogiannis and Spirakis showed that, if there is
%no pure $\frac{2}{3}$-WSNE, the min-max strategies for the zero-sum game are
%always a $\frac{2}{3}$-WSNE in the original game~\cite{KS}. 
To find an improvement over the KS algorithm we use the linear programs from
Definitions~\ref{def:rearrangey} and~\ref{def:rearrangex} with parameters $S_r =
\supp(\x)$ and $S_c = \supp(\y)$. Let $(\x^*, \y^*)$ be the mixed strategy
profile returned by the LPs, and let $\epsilon_i$ be the smallest value such
that $(\x^*, \y^*)$ is a $\epsilon_i$-WSNE.
\end{itemize}
After executing these three procedures, we take the smallest of $\epsilon_p$,
$\epsilon_m$, and $\epsilon_i$, and return the corresponding WSNE. Since all
three procedures can be implemented in polynomial time, this is a polynomial
time algorithm. The rest of this paper is dedicated to proving the following
theorem.

\begin{theorem}
\label{thm:main}
Our algorithm finds a $(\frac{2}{3} - 0.005913759)$-WSNE.
\end{theorem}

\section{Proof Outline}
\label{sec:roadmap}

In order for our proof to be as informative as possible, we will parameterize it
using a constant $z > 0$. We will show the conditions under which our algorithm
can produce a $(\frac{2}{3} - z)$-WSNE. At the end of the proof we will show
that these conditions are satisfied for $z = 0.005913759$, which provides a
proof Theorem~\ref{thm:main}.

Our approach is to assume that Procedures~1 and~2 did not produce a
$(\frac{2}{3} - z)$-WSNE, and then to use that assumption to determine the
conditions under which Procedure~3 does find a $(\frac{2}{3} - z)$-WSNE.
This comprises of the following steps.
\begin{itemize}
\item \textbf{Reanalyze the KS algorithm.} The original analysis for the KS
algorithm assumed that the game does not have a pure $\frac{2}{3}$-WSNE.
However, in our analysis, we have assumed only that Procedure~1 did not find a
pure $(\frac{2}{3} - z)$-WSNE, so the original KS analysis is no longer valid.
In Section~\ref{sec:reks} we show that, assuming there is no pure $(\frac{2}{3}
- z)$-WSNE, the KS algorithm will produce a strategy profile $(\x, \y)$ where
all strategies have payoff at most $\frac{2}{3} + 2z$, and therefore $(\x, \y)$
is a $(\frac{2}{3} + 2z)$-WSNE.
\item \textbf{Study the bad strategies.} Our goal is to show that $(\x, \y)$
can be improved from a $(\frac{2}{3} + 2z)$-WSNE to a $(\frac{2}{3} - z)$-WSNE.
To achieve this, we show how to reduce the payoffs of all strategies from
$\frac{2}{3} + 2z$ to $\frac{2}{3} - z$. While describing our approach, we will
focus on the row player, but all of our techniques will actually be applied to
both players.
We define a \emph{bad} row to be a row that has payoff strictly more
than $\frac{2}{3} - z$. 
In Section~\ref{sec:structure} we study the properties of bad rows, and
we prove that all bad rows are similar in structure to
the games shown in Figures~\ref{fig:ex1} and~\ref{fig:ex2}. That is, most of the columns in
a bad row are either \emph{big} (ie. close to $1$,) or \emph{small} (ie.
close to $\frac{1}{3}$.) We prove lower bounds on the amount of probability that
the column player's strategy assigns to big and small payoffs. We then define a
new strategy for the column player $\yimp$, that finds the \emph{worst} bad row
$\rowb$ (ie. the row with the largest payoff,) and shifts all probability from
the big columns in $\rowb$ to the small columns in $\rowb$. 
\item \textbf{Apply the matching pennies argument.} In Section~\ref{sec:mp} we
use the fact that Procedure 2 did not find an $\epsilon$-WSNE on a $2 \times 2$
support with $\epsilon < \frac{2}{3} - z$. Intuitively, this corresponds to
ruling out cases like the one shown in Figure~\ref{fig:ex2}. We prove that, if
Procedure 2 failed to find a $(\frac{2}{3} - z)$-WSNE, then the bad rows cannot
be arranged like they are in Figure~\ref{fig:ex2}. This gives a formal condition
on how the probability of $\y$ can be distributed over the columns of the bad
rows, which will be used later in the proof.
\item \textbf{Find an improved strategy.} Since the strategy shifts all
probability from big payoffs in row $\rowb$ to small payoffs in row $\rowb$, by
definition, we must have that the payoff of $\rowb$ against $\yimp$ is small.
However, the payoff of other rows may increase as we move from $\y$ and $\yimp$.
We must find a trade-off between the bad rows decreasing in payoff, and other
rows increasing in payoff, so we define a strategy $\y(t) = (1 - t) \cdot \y + t
\cdot \yimp$, which mixes between $\y$ and $\yimp$. We show that there exists a
$t$ such that all rows $i$ have payoff less than or equal to $\frac{2}{3} - z$
against $\y(t)$. In Section~\ref{sec:new} we develop a computer assisted proof
for this task. For each $z$ and $t$ we formulate a linear program that gives the
largest possible payoff of a row against $\y(t)$, and then we perform a grid
search over $z$ and $t$ in order to find a strategy $\y(t)$ against which all
rows have payoff at most $\frac{2}{3} - z$. Ultimately, we find that this occurs
for $z = 0.005913759$, which proves Theorem~\ref{thm:main}.
\end{itemize}

Before we continue with the proof, we justify why it is possible to treat the
two players independently in our analysis. In our proof, we will start with a
strategy profile $(\x, \y)$. At a high level, the idea is to rearrange the
probabilities in $\x$ to create $\x'$ such that the column player happier when
he plays $\y$ against $\x'$. Simultaneously, we rearrange the probabilities in
$\y$ to create $\y'$ such that the row player is happier when he plays $\x$
against $\y'$. We then claim that both players are happier in the profile $(\x',
\y')$. To see why, observe that an approximate well supported Nash equilibrium
is defined entirely by the supports that the strategies use. Since we only
rearrange probabilities, ie.\ we have $\supp(\x') \subseteq \supp(\x)$ and
$\supp(\y') \subseteq \supp(\y)$, it is sufficient to consider only $\x'$
played against $\y$ and $\y'$ played against $\x$ in order to prove properties
of $(\x', \y')$.

\section{Reanalyzing the KS algorithm}
\label{sec:reks}

%The original KS algorithm uses a preprocessing step that checks for a {\em pure}
%$\frac{2}{3}$-WSNE, and stops if one is found.
%In our version we initially check for a pure $\frac{2}{3}-z$-WSNE,
%a stronger requirement that leaves more input games that have to be
%handled by the rest of the algorithm. The results we establish for the
%rest of the algorithm are given in terms of the column player's strategy;
%corresponding results hold when the row player is considered.

%\begin{proposition}
%\label{prop:bad}
%Assume that $\epsilon_p > \frac{2}{3} - z$, and let $(\x, \y)$ be the WSNE
%returned by the KS algorithm. If the row player has regret larger than
%$\frac{2}{3}-z$ in $(\x, \y)$, then  for all rows $i'$ we have both of the following:
%\begin{align*}
%R_{\rowx} \cdot \y \le \frac{2}{3} + 2z, && R_{\rowx} \cdot \y - C_{\rowx}
%\cdot \y \leq 3z.
%\end{align*}
%\end{proposition}

In this section we analyse the KS algorithm 
under the assumption that Procedure~1  did not find
a $(\frac{2}{3} - z)$-WSNE. 
%We start by applying the assumption that $\epsilon_p
%> \frac{2}{3} - z$. 
Note that if there is a pure strategy profile $(\row,\col)$,
such that $R_{\row,\col} \ge \frac{1}{3} + z$ and $C_{\row,\col} \ge \frac{1}{3}
+ z$, then $(\row, \col)$ is a $(\frac{2}{3} - z)$-WSNE. 
Therefore, our assumption allows us to conclude that 
for all $\row$ and $\col$ we have:
%there cannot be a pair of pure strategies with that property. Since all payoffs
%in $R$ and $C$ lie in the range $[0, 1]$, this implies, 
\begin{equation}\label{eqn:sum}
0 \le R_{\row,\col} + C_{\row,\col} \le \frac{4}{3} + z.
\end{equation}
This inequality replaces the inequality $0 \le R_{\row,\col} + C_{\row,\col} \le
\frac{4}{3}$, which was used in the original analysis. 

From now on, our analysis will be stated for the row player, with the
understanding that all of our proofs can be apply symmetrically to the column
player. Our goal is to show that all ``worst-case'' examples for the KS algorithm
are similar to Figures~\ref{fig:ex1} and~\ref{fig:ex2}. More precisely, if
$(\x, \y)$ is the strategy profile returned by the KS algorithm, then we are
interested in 
the following properties of Figures~\ref{fig:ex1} and~\ref{fig:ex2}:
\begin{itemize}
\item There exists a row $i \in \supp(\x)$ such that $R_i \cdot \y = 0$
and $C_i \cdot \y = 0$.
\item Every row $i$ with $R_i \cdot \y = \frac{2}{3}$ also has $C_i
\cdot \y = \frac{2}{3}$.
\end{itemize}
We will show that our ``worst-case'' examples have similar properties.
%Recall that, in order to find a WSNE, the KS algorithm solves a zero-sum game
%$(D, -D)$ where $D = \frac{1}{2}(R - C)$. Suppose that we solve the game, and
%that we obtain a mixed strategy profile $(\x, \y)$. If $(\x, \y)$ happens to be
%a $(\frac{2}{3} - z)$-WSNE, then our proof is complete. Otherwise, at least one
%of the players has regret larger than $\frac{2}{3} - z$. We will suppose that
%this is the row player, and we will provide proofs for this scenario. However,
%all of our techniques can be applied symmetrically to the other case.

%Recall the worst-case example that was presented in Figure~\ref{fig:ex1}.
%There we saw an instance where the row player had regret $\frac{2}{3}$, because
%there was a row in the support with payoff~$0$, and a row outside the support
%with payoff $\frac{2}{3}$. We will show that, if the row player has regret
%larger than $\frac{2}{3} - z$ in $(\x, \y)$, then the game must necessarily be
%similar to the example of Figure~\ref{fig:ex1}. We begin by showing that there
%must be a row in the support of $\x$ with payoff close to $0$.

We begin with the first property. Here we show that, if $(\x, \y)$ is not a
$(\frac{2}{3} - z)$-WSNE, then there exists a row in the row player's support
where both players have payoff close to $0$.

\begin{proposition}
\label{prop:x}
If $(\x, \y)$ is a solution of $(D, -D)$ such that there is an $i \in \supp(\x)$
where $i$ is not a $(\frac{2}{3} - z)$-best response against $\y$ in $(R, C)$,
then there is a row $\row \in \supp(\x)$ such that both of the following hold:
\begin{align*}
R_{\row} \cdot \y < 3z, && C_{\row} \cdot \y < 3z.
\end{align*}
\end{proposition}
\begin{proof}
We begin by noting that, since $D = \frac{1}{2}(R - C)$, if $X = -\frac{1}{2}(R
+ C)$, then we have two equalities:
\begin{align*}
R = D - X, && C = -D - X.
\end{align*}
Since $\x$ is a min-max strategy in $(D, -D)$, if $\row$ is a row in
$\supp(\x)$, then for all rows $\rowx$ we have:
\begin{align*}
D_{\row} \cdot \y &\geq D_{\rowx} \cdot \y, \\
(R+X)_{\row} \cdot \y &\geq (R+X)_{\rowx} \cdot \y, \\
R_{\row} \cdot \y &\geq R_{\rowx} \cdot \y - (X_{\row} - X_{\rowx}) \cdot \y.
\end{align*}

%Since the row player has regret larger than $\frac{2}{3} - z$, when $(\x, \y)$
%is played in $(R, C)$, there must be a pair of rows 
Let $\row \in \supp(\x)$ be a row that is not a $(\frac{2}{3} -
z)$-best response against $\y$, which exists by assumption, and let $\rowx$ be a
best-response against $\y$. We have:
 %there must be a pair of rows
%$\row, \rowx$ with $\row \in
%\supp(\x)$, and $\rowx \notin \supp(\x)$ such that:
\begin{align*}
R_{\rowx} \cdot \y - (\frac{2}{3} - z) &> R_{\row} \cdot \y, \\
&\ge R_{\rowx} \cdot \y - (X_{\row} - X_{\rowx}) \cdot \y.
\end{align*}
Hence, we have:
\begin{equation*}
(X_{\row} - X_{\rowx}) \cdot \y > (\frac{2}{3} - z).
\end{equation*}
Note that, by Equation~\eqref{eqn:sum}, all
entries of $X$ must lie in the range $[-\frac{2}{3} - \frac{1}{2}z, 0]$. In
particular, this implies that:
\begin{equation*}
-\frac{2}{3} - \frac{1}{2}z \le X_{\rowx} \cdot \y < X_{\row} \cdot \y -
(\frac{2}{3} - z).
\end{equation*}
This implies that $-\frac{3}{2}z < X_\row \cdot \y \le 0$. Now, using the
definition of $X$ we obtain:
\begin{equation*}
-\frac{1}{2}(R + C)_\row \cdot \y > -\frac{3}{2}z,
\end{equation*}
which is equivalent to:
\begin{equation*}
(R + C)_\row \cdot \y < 3z.
\end{equation*}
Since both $R$ and $C$ are non-negative, we have completed the proof. 
\end{proof}

%The other feature of the example given in Figure~\ref{fig:ex1} is that
%there is a row $\rowx \notin \supp(\x)$ in which both $R_\rowx \cdot \y =
%\frac{2}{3}$ and $C_\rowx \cdot \y = \frac{2}{3}$. The next proposition shows
%that, whenever the algorithm produces a strategy profile that is not a
%$(\frac{2}{3} - z)$-WSNE, then such a row must always exist. We prove this by
%showing that $R_{\rowx} \cdot \y - C_{\rowx} \cdot \y \leq 3z$ holds for all
%rows $\rowx$.

We now consider the second property. Here we show that, if $(\x, \y)$ is not a
$(\frac{2}{3}-z)$-WSNE, then every row has payoff at most $\frac{2}{3} + 2z$,
and that for all rows $i$ we have that $R_\row \cdot \y - C_\row \cdot \y$ is
small.

\begin{proposition}
\label{prop:bad2}
If $(\x, \y)$ is a solution of $(D, -D)$ 
such that there is an $i \in \supp(\x)$
where $i$ is not a $(\frac{2}{3} - z)$-best response against $\y$ in $(R, C)$,
%such that the row player has regret
%larger than $\frac{2}{3}-z$ when $(\x, \y)$ is played in  $(R, C)$, 
then for all
rows $\rowx$ both of the following hold:
\begin{align*}
R_{\rowx} \cdot \y \le \frac{2}{3} + 2z, && R_{\rowx} \cdot \y - C_{\rowx}
\cdot \y \leq 3z.
\end{align*}
\end{proposition}
\begin{proof}
Let $\row$ be the row in $\supp(\x)$ whose existence is implied by
Proposition~\ref{prop:x}. This proposition, along with the fact that all entries
in $R$ and $C$ are non-negative, implies that:
\begin{align*}
0 \le R_\row \cdot \y < 3z, && 0 \le C_\row \cdot \y < 3z.
\end{align*}
By definition we have $D = \frac{1}{2}(R - C)$, and therefore, we have:
\begin{equation*}
-\frac{3}{2}z < D_\row \cdot \y < \frac{3}{2} z.
\end{equation*}

Now, since $x$ is a min-max strategy for the zero-sum game $(D, -D)$, we must
have, for all rows $\rowx$:
\begin{equation*}
D_{\rowx} \cdot \y \le D_{\row} \cdot \y < \frac{3}{2} z.
\end{equation*}
Thus, we have:
\begin{equation*}
\frac{1}{2}(R_{\rowx} - C_{\rowx}) \cdot \y < \frac{3}{2} z.
\end{equation*}
Rearranging this yields one of our two conclusions:
\begin{equation}
\label{eqn:diffrc}
R_{\rowx} \cdot \y < C_{\rowx} \cdot \y + 3z.
\end{equation}

We can obtain the other conclusion by rearranging Equation~\eqref{eqn:sum}, to
argue that for all rows $\row$ and all columns $\col$ we have:
\begin{equation*}
C_{\row,\col} \le \frac{4}{3} + z - R_{\row,\col}.
\end{equation*}
Then, Equation~\eqref{eqn:diffrc} implies that:
\begin{equation*}
R_{\rowx} \cdot \y < C_{\rowx} \cdot \y + 3z \le \frac{4}{3} + 4z -
R_{\rowx} \cdot \y.
\end{equation*}
This implies that $2 \cdot R_{\rowx} \cdot \y \le \frac{4}{3} + 4z$, and so
we have $R_{\rowx} \cdot \y \le \frac{2}{3} + 2z$. 
\end{proof}

Proposition~\ref{prop:bad2} shows that $R_{\rowx} \cdot \y \le \frac{2}{3} + 2z$
holds for all rows $\rowx$. Using the same argument symmetrically, we can also
show that $C_{\colx} \cdot \x \le \frac{2}{3} + 2z$ for all columns $\colx$.
Thus, we have shown that if there is no pure $(\frac{2}{3} - z)$-WSNE, then the
KS algorithm will produce a mixed strategy pair $(\x, \y)$ that is a
$(\frac{2}{3} + 2z)$-WSNE. 

The main goal of our proof is to show that the probabilities  in $\x$ and $\y$
can be rearranged to construct a $(\frac{2}{3} - z)$-WSNE. From this point
onwards, we only focus on improving the strategy $\y$, with the understanding
that all of our techniques can be applied in the same way to improve the
strategy $\x$. For the rest of the paper, we will fix $(\x, \y)$ to be the
strategy profile produced by the KS algorithm, and we will assume that it is not
a $(\frac{2}{3} - z)$-WSNE.

\section{Bad Rows}
\label{sec:structure}

In order to transform $(\x, \y)$ to a $(\frac{2}{3} - z)$-WSNE, we will ensure
that there are no rows with payoff greater than $\frac{2}{3} - z$. Thus, we
define a \emph{bad} row to be a row $\row$ whose payoff lies in the range
$\frac{2}{3} - z < R_\row \cdot \y \le \frac{2}{3} + 2z$. Furthermore, we
classify the bad rows according to how bad they are.

\begin{definition}
\label{def:ravg}
A row~$\row$ is $q$-\emph{bad} if $R_{\row} \cdot \y = \frac{2}{3} + 2z - qz$.
\end{definition}
Since $(\x, \y)$ is a $(\frac{2}{3} + 2z)$-WSNE, we have that every row is
$q$-bad for some $q \ge 0$. Moreover, we are interested in improving the $q$-bad
rows with $0 \le q < 3$. In this section, we study the properties of $q$-bad
rows, and we show that they must look similar to the bad rows in
Figures~\ref{fig:ex1} and~\ref{fig:ex2}.

To begin, we observe that if $\row$ is a $q$-bad row, then we can apply the second inequality
of Proposition~\ref{prop:bad2} to obtain:
\begin{equation}
\label{eqn:cavg}
C_{\row} \cdot \y \geq \frac{2}{3} - z - qz.
\end{equation}
Now consider a~$q$-bad row~$\row$ with $q < 3$. We can deduce the following
three properties about row~$\row$.
\begin{itemize}
\item Definition~\ref{def:ravg} tells us that 
$R_{\row} \cdot \y$ is close to $\frac{2}{3}$.
\item Equation~\eqref{eqn:cavg} tells us that 
$C_{\row} \cdot \y$ is close to $\frac{2}{3}$.
\item The fact that there are no pure $(\frac{2}{3} - z)$-WSNEs implies that,
for each column $\col$, we must either have $R_{\row, \col} < \frac{1}{3} + z$
or $C_{\row, \col} < \frac{1}{3} + z$, because otherwise $(\row, \col)$ would be
a pure $(\frac{2}{3} - z)$-WSNE.
\end{itemize}
In order to satisfy all three of these conditions simultaneously, the row $\row$
must have a very particular form: approximately half of the probability assigned
by $\y$ must be given to columns~$\col$ where $R_{\row, \col}$ is close to~$1$
and~$C_{\row, \col}$ is close to~$\frac{1}{3}$, and approximately half of the
probability assigned by $\y$ must be given to columns~$\col$ where $R_{\row,
\col}$ is close to $\frac{1}{3}$ and $C_{\row, \col}$ is close to~$1$.

Building on this observation, we split the columns of each row $\row$ into three
sets. We define the set~$B_{\row}$ of \emph{big} columns to be $B_{\row} = \{
\col \; : \; R_{\row,\col} \ge \frac{2}{3} + 2z \}$, and the set~$S_{\row}$ of
\emph{small} columns to be $S_{\row} = \{ \col \; : \; C_{\row,\col} \ge
\frac{2}{3} + 2z \}$. Finally, we have the set of \emph{other} columns $O_{\row}
= \{1,2,\dots,n\} \setminus (B_{\row} \cup S_{\row})$, which contains all
columns that are neither big nor small. 

We now formalise our observations by giving inequalities about the amount of
probability that $\y$ can assign to $B_\row$, $S_\row$, and $O_\row$, for every
$q$-bad row $\row$. The following proposition proves three inequalities. The
first inequality is proved using Markov's inequality. The second and third
inequalities arise from substituting the first inequality into
Definition~\ref{def:ravg} and Equation~\eqref{eqn:cavg}, respectively. The full
proof of this proposition is presented in~\ref{app:inequalities}.

\begin{proposition}
\label{prop:inequalities}
If $i$ is a $q$-bad row then:
\begin{align*}
\sum_{\col \in O_{\row}} \y_{\col} &\le \frac{2qz}{\frac{1}{3} - 2z}, \\
\sum_{\col \in B_{\row}} \y_{\col} &\ge \frac{\frac{1}{3} + z -
qz - (\frac{1}{3} + z) \sum_{\col \in O_{\row}} \y_{\col}}{\frac{2}{3} - z}, \\
\sum_{\col\in S_{\row}} \y_{\col} &\ge \frac{\frac{1}{3} -2z-qz
- (\frac{1}{3} + z)\sum_{\col\in O_{\row}} \y_{\col}}{\frac{2}{3} - z}.
\end{align*}
\end{proposition}

\paragraph{\bf The strategies $\yimp$ and $\y(t)$}

We now define our improved strategies. Let $\rowb$ to be a \emph{worst} bad row.
That is $\rowb$ is a row that satisfies $\arg\max_{\row}(R_{\row} \cdot \y)$,
and therefore ${\rowb}$ is a $\qb$-bad row such that there is no~$q$-bad row
with $q<\qb$. We fix~$\rowb$ and~$\qb$ to be these choices for the rest of this
paper. Note that we can assume that $\qb < 3$, because if this is not the case,
then all rows have payoff less than or equal to $\frac{2}{3} - z$, and $\y$ does
not need to be improved. 

We begin by defining a strategy that improve row $\rowb$.
We will improve row~${\rowb}$ by moving the probability assigned to
$B_{\rowb}$ to $S_{\rowb}$. 
%This is a generalisation of shifting probability
%from the first column to the second column in Figure~\ref{fig:ex1}. 
Formally, we
define the strategy $\yimp$ as follows. For each~$\col$ with $1 \le \col \le n$,
we have:
\begin{equation}
\label{def:yimp}
\yimp_{\col} = \begin{cases}
0 & \text{if $\col \in B_{\rowb}$}, \\
\y_{\col} + \frac{\y_{\col} \cdot \sum_{k \in B_{\rowb}} \y_{k}}{\sum_{k\in
S_{\rowb}} \y_{k}}
  & \text{if $\col \in S_{\rowb}$}, \\
\y_{\col} & \text{otherwise.}
\end{cases}
\end{equation}

The strategy $\yimp$ improves the specific bad row $\rowb$, but other rows may
not improve, or even get worse in $\yimp$. Therefore, we will study convex
combinations of $\y$ and $\yimp$. More formally, for the parameter $t \in [0,
1]$, we define the strategy $\y(t)$ to be $(1-t) \cdot \y + t \cdot \yimp$.

\section{Applying the matching pennies argument}
\label{sec:mp}

%\todo[inline]{Re-merge this section}
%\todo[inline]{Talk about matching pennies subgames in $\y$}

%Recall that $\epsilon_m$ is computed in stage 2 of our algorithm, and
%is the quality of the best WSNE with $2\times 2$ support.
So far, we have not used the assumption that
Procedure 2 did not find a $(\frac{2}{3} - z)$-WSNE.
 %$\epsilon_m > \frac{2}{3} - z$.
In this section we will see how this assumption can be used to prove properties
about the $q$-bad rows.
We begin by defining the concept of a matching pennies sub-game.

\begin{definition}[Matching Pennies]
\label{def:mp}
Let $\y$ be a column player strategy, let
$\row$ and $\rowx$ be two rows, and let $\col$ and $\colx$ be two columns.
If $\col \in B_\row \cap S_{\rowx}$ and $\colx \in B_{\rowx} \cap S_\row$, then
we say that $\row$, $\rowx$, $\col$, and $\colx$ form a matching pennies
sub-game in $\y$.
\end{definition}
An example of a matching pennies sub-game is given by $l$, $r$, $T$, and $M$ in
Figure~\ref{fig:ex2}, because we have $l \in B_M \cap S_T$, and we have $r
\in B_T \cap S_M$. In this example, we can obtain an exact Nash equilibrium by
making the row player mix uniformly between~$T$ and~$M$, and making the column
player mix uniformly between~$l$ and~$r$. However, in general we can only expect
to obtain an $(\frac{2}{3} - z)$-WSNE using this technique, as the following
proposition shows.

\begin{proposition}
\label{prop:mp}
Let $\y$ be a column player strategy. If there is a matching pennies sub-game in
$\y$, then we can construct a $(\frac{2}{3} - z)$-WSNE with a $2 \times 2$
support.
\end{proposition}
\begin{proof}
Let $\row$, $\rowx$, $\col$, and $\colx$ be a matching pennies sub-game in
$\y$. We define two strategies $\x'$ and $\y'$ as follows:
\begin{align*}
\x'_k = \begin{cases}
0.5 & \text{if $k = \row$ or $k = \rowx$}, \\
0 & \text{otherwise.}
\end{cases}
&&
\y'_k = \begin{cases}
0.5 & \text{if $k = \col$ or $k = \colx$}, \\
0 & \text{otherwise.}
\end{cases}
\end{align*}
We will prove that $(\x', \y')$ is a $(\frac{2}{3} - z)$-WSNE. Note that when
the column player plays $\y'$, the payoff to the row player from row $\row$ is:
\begin{equation*}
R_{\row} \cdot \y' = 0.5 \cdot R_{\row,\col} + 0.5 \cdot R_{\row,\colx}.
\end{equation*}
Since $\col \in B_{\row}$ we have $R_{\row, \col} \ge \frac{2}{3} + 2z$, Hence, we have:
\begin{equation*}
R_{\row} \cdot \y' \ge 0.5 \times \left( \frac{2}{3} + 2z \right) + 0.5 \times 0 = \frac{1}{3} + z.
\end{equation*}
An identical argument can be used to argue that $R_{\rowx} \cdot \y'$, ${C^T}_{\col} \cdot
\x'$, and ${C^T}_{\colx} \cdot \x'$ are all greater than or equal to $\frac{1}{3} + z$.

Thus, we have shown that all pure strategies in the support of $\x'$ and $\y'$
are $(\frac{2}{3} - z)$-best responses. 
%Since $R_{k} \cdot \y' \le 1$ and $C_k \cdot \x' \le 1$ for all $k$, the largest
%possible regret that can be experienced by either of the two players is $1 -
%(\frac{1}{3} + z) = \frac{2}{3} - z$. 
Hence, $(\x', \y')$ is a $(\frac{2}{3} - z)$-WSNE. 
\end{proof}

Proposition~\ref{prop:mp} allows us to assume that the game does not contain a
matching pennies sub-game, because otherwise Procedure 2 would have found a
$(\frac{2}{3} - z)$-WSNE. Note that, by definition, if the game does not contain
a matching pennies sub-game, then for all rows $\row$ we must have either
$B_\rowb \cap S_\row = \emptyset$, or $B_\row \cap S_\rowb = \emptyset$.

%We can use this observation to strengthen our LP. We define two LPs, each of
%which is constructed by adding an extra constraint to our existing LP. In the
%first LP we add the constraint $\var_{bs} = 0$, and in the second LP we add the
%constraint $\var_{sb} = 0$. We refer to the solutions of these two LPs as
%$\sol_1(z, \qb, q)$ and $\sol_2(z, \qb, q)$ respectively. We then obtain the
%following strengthening of Proposition~\ref{prop:allq}.

%\begin{proposition}
%\label{prop:mplp}
%For each $q$-bad row $\row$ we either
%have $R_\row \cdot \yimp \le \sol_1(z, \qb, q)$, or we have $R_\row \cdot \yimp
%\le \sol_2(z, \qb, q)$.
%\end{proposition}

\section{An Improved Strategy Exists}
\label{sec:new}

Our goal is to show that there exists a $t$ in the range $0 \le t \le 1$ and a
$z>0$ such that for every row $i$, we have $R_i \cdot \y(t) \le \frac{2}{3} -
z$. In this section, we develop a computer assisted proof of this fact.

Recall that the strategy $\yimp$ is defined by moving all probability from the
columns in $B_{\rowb}$ to the columns in $S_{\rowb}$. We are interested in how
other rows $\row$ are affected by this operation. This will depend on how much
probability mass is shared between the partition $(B_i, S_i, O_i)$, and the
partition $(B_{\rowb}, S_{\rowb}, O_{\rowb})$. Figure~\ref{fig:intersect}
shows the nine possible intersections.

%By definition, since all
%row-player payoffs in $S_{\rowb}$ are smaller than the row-player payoffs in
%$B_{\rowb}$, this can only reduce the payoff of row $\rowb$. Thus, as $t$
%is increases away from $0$, we have that $R_{\rowb} \cdot \y(t)$ decreases.

%However, in order to show that \emph{all} rows $i$ satisfy $R_i \cdot \y(t) \le
%\frac{2}{3} - z$, we must understand how the rows $i$ with $i \ne \rowb$ behave
%as $t$ is increases away from $0$. 

\begin{figure}[h]
\begin{center}
\begin{tikzpicture}[scale=0.6]
\draw (13.5,1) rectangle (0,0);
\draw (13.5,3) rectangle (0,2);
\node at (-1, 0.5) {Row $\row$};
\node at (-1, 2.5) {Row $\rowb$};
\draw (13.5/3,2) -- (13.5/3,3);
\draw (13.5*2/3,2) -- (13.5*2/3,3);
\node at (13.5/6,2.5) {$B_{\rowb}$};
\node at (13.5/3 + 13.5/6,2.5) {$S_{\rowb}$};
\node at (13.5*2/3 + 13.5/6,2.5) {$O_{\rowb}$};
\foreach \x in {1,...,8}
	\draw (13.5*\x/9,0) -- (13.5*\x/9,1);
\foreach \x in {0,3,6}
	\node at (13.5 * \x/9 + 13.5/18,0.5) {$B_{\row}$};
\foreach \x in {1,4,7}
	\node at (13.5 * \x/9 + 13.5/18,0.5) {$S_{\row}$};
\foreach \x in {2,5,8}
	\node at (13.5 * \x/9 + 13.5/18,0.5) {$O_{\row}$};
\end{tikzpicture}
\end{center}
\caption{The nine possible intersections between the partition $(B_{\rowb},
S_{\rowb}, O_{\rowb})$, and the partition $(B_{\row}, S_{\row}, O_{\row})$ for the rows $i$
and $\rowb$.}
\label{fig:intersect}
\end{figure}

We are interested in the amount of probability that $\y$ assigns to each of
these nine sets. We define a shorthand for this purpose:
\begin{align*}
\dbb &= \sum\nolimits_{j \in B_\rowb \cap B_{\row}} \y_j, & \dsb &= \sum\nolimits_{j \in S_\rowb \cap
B_{\row}} \y_j, & \dob &= \sum\nolimits_{j \in O_\rowb \cap B_{\row}} \y_j, \\
\dbs &= \sum\nolimits_{j \in B_\rowb \cap S_{\row}} \y_j, & \dss &= \sum\nolimits_{j \in S_\rowb \cap
S_{\row}} \y_j, & \dos &= \sum\nolimits_{j \in O_\rowb \cap S_{\row}} \y_j, \\
\dbo &= \sum\nolimits_{j \in B_\rowb \cap O_{\row}} \y_j, & \dso &= \sum\nolimits_{j \in S_\rowb \cap
O_{\row}} \y_j, & \doo &= \sum\nolimits_{j \in O_\rowb \cap O_{\row}} \y_j. \\
\end{align*}
As $t$ is increased away from $0$, probability will be shifted from $\dbb$,
$\dbs$, and $\dbo$ to $\dsb$, $\dss$, and $\dso$, while the amount of
probability assigned to $\dob$, $\dos$, and $\doo$ will remain constant.

For each fixed $t$ in the range $0 \le t \le 1$, and each fixed $z > 0$,  we are
interested in the worst-case value of $R_i \cdot \y(t)$. We will show that an
upper bound on $R_i \cdot \y(t)$ can be obtained by solving a linear program.
The linear program has eleven variables. We use nine variables, $\dbb$, $\dbs$,
$\dbo$, $\dsb$, $\dss$, $\dso$, $\dob$, $\dos$, and $\doo$, to represent the
amount of probability assigned to the columns in $i$.  We use two additional
variables $\dq$ and $\dqb$ to represent how bad the rows $\rowb$ and $\row$ are.
These two variables should be interpreted as follows: row $\rowb$ is a
$\dqb$-bad row and row $\row$ is a $\dq$-bad row.

We can now define the linear program. We begin by defining a helper function
$\phi(z, \qb)$ as follows:
\begin{equation*}
\phi(z, \qb) = 
1  + \frac{\frac{1}{3} + z + \qb z 
}{
\frac{1}{3} -2z-\qb z - (\frac{1}{3} + z) \cdot \frac{2\qb z}{\frac{1}{3} - 2z}
}.
\end{equation*}
Our linear program will be parameterised: for each $z$ with $z \ge 0$, each $t$
in the range $0 \le t \le 1$, and each $k \in \{0, 1\}$ we define $\lp(z, t, k)$
to be the linear program shown in Figure~\ref{fig:lp}. The rest of this section
is dedicated to showing that this linear program can be used to find an upper
bound on $R_{i} \cdot \y(t)$, for all rows $i$.

\begin{figure}
\begin{tcolorbox}
\noindent Maximize:
\begin{multline*}
(1 - t) \Biggl(\frac{2}{3} + 2 z - \dq \cdot z\Biggr) + t \cdot \Biggl(
\phi(z, 3) \Bigl(\dsb +
		(\frac{1}{3} + z) \cdot \dss  + (\frac{2}{3} + 2 z) \cdot \dso \Bigr)
\\ + \dob +
(\frac{1}{3} + z) \cdot \dos + (\frac{2}{3} + 2z) \cdot \doo \Biggr)
\end{multline*}
Subject to:
\begin{align}
% First
\label{const:1}
\dbb + \dbs + \dbo &\ge \frac{\frac{1}{3} + z - \dqb \cdot z - (\frac{1}{3} + z)(\dob +
		\dos + \doo)}{\frac{2}{3} - z} \\
% Second
\label{const:2}
\dbb + \dsb + \dob &\ge \frac{\frac{1}{3} + z - \dq \cdot z - (\frac{1}{3} + z) (\dbo
		+ \dso + \doo)}{\frac{2}{3} - z} \\
% Third
\label{const:3}
\dsb + \dss + \dso &\ge \frac{\frac{1}{3} - 2 z - \dqb \cdot z -
(\frac{1}{3} + z)(\dob + \dos + \doo)}{\frac{2}{3} - z} \\
% Fourth
\label{const:4}
\dbs + \dss + \dos &\ge \frac{\frac{1}{3} - 2 z - \dq \cdot z -
(\frac{1}{3} + z)(\dbo + \dso + \doo)}{\frac{2}{3} - z} \\
% Fifth
\label{const:5}
\dob + \dos + \doo &\le \frac{2 \cdot \dqb \cdot z}{\frac{1}{3} - 2z} \\
% Sixth
\label{const:6}
\dbo + \dso + \doo &\le \frac{2 \cdot \dq \cdot z}{\frac{1}{3} - 2z} \\
% Seventh
\label{const:7}
0 & = \begin{cases}
\dbs & \text{if $k = 0$} \\
\dsb & \text{if $k = 1$}
\end{cases}\\
\label{const:8}
\dqb & \le 3 \\
\label{const:9}
\dqb & \le \dq \\
\label{const:10}
1 &= \dbb + \dbs + \dbo + \dsb + \dss + \dso + \dob + \dos + \doo \\
\label{const:11}
0 &\le \dbb, \dbs, \dbo, \dsb, \dss, \dso, \dob, \dos, \doo, \dq, \dqb
\end{align}
\end{tcolorbox}
\caption{The linear program $\lp(z, t, k)$.}
\label{fig:lp}
\end{figure}

\paragraph{\bf The Constraints} We begin by arguing that all of the constraints
in the LP are valid. Firstly, since $z$ and $t$ are both constants, it can be
seen that all of the constraints are indeed linear. Constraints~\eqref{const:1}
through~\eqref{const:6} are taken directly from
Proposition~\ref{prop:inequalities}. Each inequality in
Proposition~\ref{prop:inequalities} appears twice: once for the row $\row$ and
once for the row $\rowb$. 

Constraint~\eqref{const:7} encodes the matching pennies argument. By
Proposition~\ref{prop:mp} if we have both $\dbs > 0$ and $\dsb > 0$, then we can
find a $(\frac{2}{3} - z)$-WSNE. Thus, we can assume that either $\dsb = 0$ or
$\dsb = 0$. Constraint~\eqref{const:7} encodes this using the parameter $k$: if
$k = 0$ then $\dbs$ is constrained to be $0$, and if $k = 1$, then $\dsb$ is
constrained to be $0$.

Constraints~\eqref{const:8} and~\eqref{const:9} provide bounds for $\dq$ and
$\dqb$. Recall that a row $\row$ is $q$-bad if $R_\row \cdot \y = \frac{2}{3} +
2 z - qz$. Since $\dqb$ is the $q$-value for a \emph{worst} bad row, and since a
worst bad row $\rowb$ must have $R_{\rowb} \cdot \y \ge \frac{2}{3} - z$, we
must have $\dqb \le 3$. This is encoded in Constraint~\eqref{const:8}.
Constraint~\eqref{const:9} again uses the fact that $\dqb$ is the $q$-value of a
worst bad row: the $q$ value for every other row must be greater than or equal
to $\dqb$.

Finally, Constraints~\eqref{const:10} and~\eqref{const:11} specify that the nine
variables must be a probability distribution. They also specify that both $\dq$
and $\dqb$ must be non-negative, which is valid because
Proposition~\ref{prop:bad2} implies that no row $\row$ can have $R_i \cdot \y >
\frac{2}{3} + 2z$.

%In conclusion, every feasible solution of the linear program $\lp(z, t, k)$
%specifies a probability distribution $\y$ and two rows $\row$ and $\rowb$ such
%that $\row$ is a $\dq$-bad row and $\rowb$ is a $\dqb$-bad row. Moreover, since
%all of our constraints have arisen from properties about $q$-bad rows that we
%have shown earlier, we have that every possible 

\paragraph{\bf The Objective}
We now show that the objective function of the linear program provides an upper
bound on $R_\row \cdot \y(t)$. To prove this, we first observe that by
definition we have 
\begin{equation}
\label{eqn:ytupper}
R_\row \cdot \y(t) = (1 - t) \cdot R_\row \cdot \y + t \cdot R_\row \cdot \yimp.
\end{equation}
Since $\row$ is a $\dq$-bad row, we have that $R_\row \cdot \y = \frac{2}{3} +
2z - \dq z$. In the following proposition, we show an upper bound for $R_\row
\cdot \yimp$.

\begin{proposition}
\label{prop:yimpupper}
We have that $R_\row \cdot \yimp$ is less than or equal to:
\begin{multline*}
\Bigl(1 + \frac{\dbb + \dbs + \dbo}{\dsb + \dss + \dso}\Bigr) \Bigl(\dsb +
		(\frac{1}{3} + z) \cdot \dss  + (\frac{2}{3} + 2 z) \cdot \dso \Bigr)
\\ + \dob +
(\frac{1}{3} + z) \cdot \dos + (\frac{2}{3} + 2z) \cdot \doo.
\end{multline*}
\end{proposition}
\begin{proof}
Since $\yimp$ is obtained from $\y$ by shifting all probability from $B_{\rowb}$
to $S_{\rowb}$, we have that:
\begin{align}
\nonumber
R_{\row} \cdot \yimp 
&= \sum_{\col \in S_{\rowb}} R_{\row, \col} \cdot \yimp + \sum_{\col \in
O_{\rowb}} R_{\row, \col} \cdot \yimp \\
\nonumber
&= \sum_{\col \in S_{\rowb}} R_{\row, \col} \cdot \yimp + \sum_{\col \in
O_{\rowb}} R_{\row, \col} \cdot \y \\
\label{eqn:splitsplit}
&= 
\Bigl(1 + \frac{\dbb + \dbs + \dbo}{\dsb + \dss + \dso}\Bigr)
\sum_{\col \in S_{\rowb}} R_{\row, \col} \cdot \y + \sum_{\col \in
O_{\rowb}} R_{\row, \col} \cdot \y.
\end{align}
The second and third equalities were obtained directly from the definition of
$\yimp$ given in~\eqref{def:yimp}.

Now to obtain the claimed result we split the two sums into their constituent
parts. Firstly, we have that $\sum_{\col \in S_{\rowb}} \y = \dsb + \dss +
\dso$, and by definition we have that:
\begin{itemize}
\item $R_{i, j} \le 1$ for each $j \in
B_{\row}$,
\item $R_{i, j} \le \frac{1}{3} + z$ for each $j \in S_{\row}$, and
\item $R_{i, j} \le \frac{2}{3} + 2z$ for each $j \in O_{\row}$. 
\end{itemize}
Similarly, we split the sum $\sum_{\col \in O_{\rowb}} \y$ into $\dob + \dos +
\doo$, and apply the same bounds as above. Combining all of these bounds and
substituting them into Equation~\eqref{eqn:splitsplit} yields the claimed
result. 
\end{proof}

Substituting our two bounds into Equation~\eqref{eqn:ytupper} does give an upper
bound on $R_i \cdot \y(t)$, but this upper bound is not linear in the variables
of the linear program. To resolve this, in the next proposition we provide a
constant upper bound for one of the terms in the LP, using the auxiliary
function $\phi(z, \qb)$ that was defined earlier.

\begin{proposition}
\label{prop:imptoy}
If $z < \frac{13 - 3 \sqrt{17}}{24} \approx 0.02627$, then 
\begin{equation*}
\Bigl(1 + \frac{\dbb + \dbs + \dbo}{\dsb + \dss + \dso}\Bigr) \le \phi(z, 3).
\end{equation*}
%for all columns $\col \in S_{\rowb}$ we have $\yimp_{\col} \le 
%\y_{\col} \cdot \phi(z, \qb) \cdot$.
\end{proposition}
\begin{proof}
%By definition we have, for each $j \in S_\rowb$:
%\begin{equation*}
%\yimp_{\col} = \y_{\col} + \y_{\col} \cdot \frac{\dbb + \dbs + \dbo}
%{\dsb + \dss + \dso}
%\end{equation*}
Proposition~\ref{prop:inequalities} implies that $\dsb + \dss + \dso \ge
\frac{\frac{1}{3} - 2 z - \qb z - (\frac{1}{3} + z)(\dob + \dos +
\doo)}{\frac{2}{3} - z}$. We can apply this in order to determine the following
upper bound for $\dbb + \dbs + \dbo$.
\begin{align*}
\dbb + \dbs + \dbo &= 1 - (\dsb + \dss + \dso) - (\dob + \dos + \doo) \\
&\le 1 - \frac{\frac{1}{3} - 2 z - \qb z - (\frac{1}{3} + z)(\dob + \dos +
\doo)}{\frac{2}{3} - z} - (\dob + \dos + \doo) \\
&= 1 - \frac{\frac{1}{3} - 2 z - \qb z + (-\frac{1}{3} - z + \frac{2}{3} -
z)(\dob + \dos + \doo)}{\frac{2}{3} - z} \\
&= 1 - \frac{\frac{1}{3} - 2 z - \qb z + (\frac{1}{3} - 2z)(\dob + \dos + \doo)}{\frac{2}{3} - z} \\
&= \frac{\frac{1}{3} + z + \qb z - (\frac{1}{3} - 2z)(\dob + \dos + \doo)}{\frac{2}{3} - z} \\
&\le \frac{\frac{1}{3} + z + \qb z}{\frac{2}{3} - z}.
\end{align*}
Substituting this gives the following upper bound.
\begin{align}
\label{eqn:yimpupper}
\Bigl(1 + \frac{\dbb + \dbs + \dbo}{\dsb + \dss + \dso} \Bigr)
&\le \Bigl( 1 + \frac{\frac{1}{3} + z + \qb z 
}{
(\frac{2}{3} - z) \cdot 
(\dsb + \dss + \dso)
}\Bigr).
\end{align} 
In order to proceed we must now use a lower bound for $(\frac{2}{3} - z) \cdot
(\dsb + \dss + \dso)$. By Proposition~\ref{prop:inequalities} we have that:
\begin{align}
\nonumber
(\frac{2}{3} - z) \cdot (\dsb + \dss + \dso) 
&\ge \frac{1}{3} -2z-\qb z - (\frac{1}{3} + z)(\dob + \dos + \doo) \\
\label{eqn:yimpupper2}
&\ge \frac{1}{3} -2z-\qb z - (\frac{1}{3} + z) \cdot \frac{2\qb z}{\frac{1}{3} -
2z}.
\end{align}
In order to substitute Inequality~\eqref{eqn:yimpupper2} into
Inequality~\eqref{eqn:yimpupper}, we must have that $2z + \qb z + (\frac{1}{3} +
z) \cdot \frac{2\qb z}{\frac{1}{3} - 2z} < \frac{1}{3}$, because otherwise the
denominator of Inequality~\eqref{eqn:yimpupper} will be $0$ or negative. Since
$\qb$ can be at most $3$, this holds whenever:
\begin{align*}
5z + (\frac{1}{3} + z)\cdot \frac{6z}{\frac{1}{3} - 2z} < \frac{1}{3}.
\end{align*}
Solving this inequality for $z$ gives that $z < \frac{13 \pm 3 \sqrt{17}}{24}$.
Taking the smaller of the two solutions gives 
$z < \frac{13 - 3 \sqrt{17}}{24} \approx 0.02627$.

So, if we have $z < \frac{13 - 3 \sqrt{17}}{24}$, then we can conclude:
\begin{align*}
\Bigl(1 + \frac{\dbb + \dbs + \dbo}{\dsb + \dss + \dso} \Bigr)
& \le 
\Bigl(1 + \frac{\frac{1}{3} + z + \qb z 
}{
\frac{1}{3} -2z-\qb z - (\frac{1}{3} + z) \cdot \frac{2\qb z}{\frac{1}{3} - 2z}
} \Bigr)\\
&= \phi(z, \qb).
\end{align*}
To complete the proof we observe that, so long as $0 \le z < \frac{13 - 3
\sqrt{17}}{24}$, we have that $\phi(z, \qb)$ is monotonically increasing in
$\qb$. This holds because $\qb$ only occurs positively in the numerator and
negatively in the denominator, and because the denominator is strictly positive.
Thus, since $\qb$ can be at most $3$, we have $\phi(z, \qb) \le \phi(z, 3)$.
\end{proof}

Combining the upper bound from Proposition~\ref{prop:imptoy} with the upper
bound from Proposition~\ref{prop:yimpupper}, and substituting the result into
Equation~\eqref{eqn:ytupper} gives a linear upper bound on $R_{\row} \cdot
\y(t)$, and this linear bound is used as the objective function of the LP.

\paragraph{\bf The Upper Bound}
We can now prove that the linear program provides an upper bound on the quality
of WSNE provided by $\y(t)$. For each problem $\lp(z, t, k)$, let $\sol(\lp(z,
t, k))$ be the value of the objective function in the solution of $\lp(z, t,
k)$.

\begin{proposition}
%Let $z > 0$, let $t$ be in the range $0 \le t \le 1$, and suppose that there is
%no matching pennies sub-game in $\y$.  If $s_0$ and $s_0$ are
%the value of the objective function in the solutions of $\lp(z, t, 0)$  and
%$\lp(z, t, 1)$, respectively, then for all rows $\row$ we have:
For every $z>0$ and $t$ in the range $0 \le t \le 1$: 
\begin{itemize}
\item if $\sum_{\col \in B_{\rowb} \cap S_{\row}} \y = 0$ then
$R_{\row} \cdot \y(t) \le \sol(\lp(z, t, 0))$.
\item if $\sum_{\col \in S_{\rowb} \cap B_{\row}} \y = 0$ then
$R_{\row} \cdot \y(t) \le \sol(\lp(z, t, 1))$.
\end{itemize}
\end{proposition}
\begin{proof}
We will prove only the case where
$\sum_{\col \in B_{\rowb} \cap S_{\row}} \y = 0$, because the other case is
entirely symmetric.
Let $\row$ be a row that maximizes $R_{\row} \cdot \y(t)$, and let $\rowb$ be
the worst bad row in $\y$. It is not difficult to construct a feasible point in
$\lp(z, t, 0)$ that represents these two rows: the variables $\dbb, \dbs,
\dots$ are set according to the probability assigned to the corresponding
intersection sets by $\y$, while $\dq$ and $\dqb$ are set to be the $q$-values
of $\row$ and $\rowb$, respectively. 

Obviously, this point satisfies $\dqb \le 3$ and $\dqb \le 3$, and it also
satisfies the non-negativities and the sum-to-one constraint. Furthermore, by
assumption we have that $\dbs = 0$, so Constraint~\eqref{const:7} is satisfied.
Since all other constraints of the LP were derived from the properties of either
$\dq$-bad or $\dqb$-bad rows, we have that the point is feasible in $\lp(z, t,
0)$. 

Since Propositions~\ref{prop:yimpupper} and~\ref{prop:imptoy} show that the
objective function of the LP provides an upper bound on $R_i \cdot \y(y)$, and
since the LP is a maximization problem, we must have $R_{\row} \cdot \y(t) \le
\sol(\lp(z, t, 0))$. 
\end{proof}

\paragraph{\bf Finding $z$} We now describe how the linear programs can be used
to determine a value of $z$ such that $\y(t)$ is a $(\frac{2}{3} - z)$-WSNE. 
For every $z > 0$, if we want to prove that we can produce a $(\frac{2}{3} -
z)$-WSNE, we require a witness $(z, t_0, t_1)$ that satisfies both of the
following conditions:
\begin{itemize}
\item A $t_0$ in the range $0 \le t_0 \le 1$ such that $\sol(z, t_0, 0) \le \frac{2}{3} - z$.
\item A $t_1$ in the range $0 \le t_1 \le 1$ such that $\sol(z, t_1, 1) \le \frac{2}{3} - z$.
\end{itemize}
If a pair $(t_0, t_1)$ can be found that satisfy these properties, then
$\y(t_0)$ is a $(\frac{2}{3} - z)$-WSNE in the case where $\dbs = 0$, and
$\y(t_1)$ is a $(\frac{2}{3} - z)$-WSNE in the case where $\dsb = 0$.

Our strategy for finding a witness $(z, t_0, t_1)$ was to perform a grid search
over all possible values for $z$, $t_0$, and $t_1$ using a suitably small
increment. We implemented this approach in Mathematica, where for each candidate
witness, we solved the two linear programs in exact arithmetic. Ultimately, we
were able to find a witness $(0.005913759, 0.120, 0.168)$. We were unable to
find a witness for $z = 0.005913760$. Thus, we have completed the proof of
Theorem~\ref{thm:main}.

\section{Conclusion}

We have shown that our algorithm always finds a $(\frac{2}{3} -
0.005913759)$-WSNE. Our computer assisted proof relied upon a linear program.
We tried several ways to improve this analysis, all of which were ultimately
unsuccessful. 

The current proof finds two values of $t$: one for the case where $\dbb =
0$, and one for the case where $\dsb = 0$. One obvious approach towards
improving the analysis is to split the analysis into more cases, and compute
a $t$ for each case. One of our unsuccessful attempts in this direction
was to parameterise the LP for different values of $\dqb$. The existing LP
allows $\dqb$ to take any value in the range $[0, 3]$, but we could, for
example, use one LP for the case where $\dqb \in [0, 1.5]$ and another for the
case where $\dqb \in [1.5, 3]$, and then compute two different values of $t$ for
these two cases. Unfortunately, this did not yield a better analysis no
matter how many different bands we used.

The objective function of the LP uses a linear upper bound on the
non-linear expression from Proposition~\ref{prop:yimpupper}. We could, in
principle, attempt to solve the non-linear optimization problem that is obtained
when the expression from Proposition~\ref{prop:yimpupper} is used directly as
the objective function. Unfortunately, it seems that this task is beyond current
technology. In particular, the need to solve the problem in exact arithmetic
thwarted our attempts to solve the problem within a reasonable running time.

% Our stuff may not be tight
%In Section~\ref{sec:ouralgorithm}, we presented a polynomial-time algorithm 
%for computing a $(\frac{2}{3}-z)$-WSNE, where $z = 0.004735$.
%We do not believe that our analysis is tight, as it uses several restrictions
%that our algorithm does not face.
%For example, $\y(t)$ uses the same support as the strategy returned by the KS
%algorithm, whereas the LP given in Definition~\ref{def:rearrangey} can return
%a subset of this support.
%Another example is that in the analysis we only consider $2
%\times 2$ subgames in which players mix uniformly, whereas Procedure 2 considers
%all mixtures.

%An interesting open question is the following. 
%Does every bimatrix game possess a $\frac{1}{2}$-WSNE, where both players
%use at most two strategies?
%This is known to be true with high probability in random games~\cite{BVV07},
%but not known in general.
%\todo[inline]{This was resolved in the negative by WINE paper; would be nice to still
%mention somehow and cite the WINE paper}

\bibliographystyle{spmpsci}
\bibliography{references.bib}

\appendix
\newpage

\section{Proof of Proposition~\ref{prop:inequalities}}
\label{app:inequalities}

We begin by proving the inequality for $O_\row$. The first thing that we note is
that, if a column $\col$ is in $O_\row$, then $R_{\row, \col} + C_{\row, \col}$
must be significantly smaller than $\frac{4}{3} + z$.

\begin{proposition}
\label{prop:farsum}
For each row~${\row}$, and each column ${\col}\in O_{\row}$, we have
$R_{\row,\col} + C_{\row,\col} < 1 + 3z$.
\end{proposition}
\begin{proof}
For each column $\col \in O_{\row}$ we have both of the following properties:
\begin{itemize}
\item Since $\col \notin B_{\row}$, we have $R_{\row,\col} < \frac{2}{3} + 2z$.
\item Since $\col \notin S_{\row}$, we have $C_{\row,\col} < \frac{2}{3} + 2z$.
\end{itemize}
Furthermore, our assumption that Procedure (1) does not find a pure
$(\frac{2}{3} - z)$-WSNE implies that:
\begin{itemize}
\item If $R_{\row,\col} \ge \frac{1}{3} + z$, then $C_{\row,\col} < \frac{1}{3} + z$.
\item If $C_{\row,\col} \ge \frac{1}{3} + z$, then $R_{\row,\col} < \frac{1}{3} + z$.
\end{itemize}
This is because, if these inequalities did not hold for some pair $\row$ and
$\col$, then it is easy to show that $(\row, \col)$ is a $(\frac{2}{3} -
z)$-WSNE. From these properties it is easy to see that $R_{\row,\col} +
C_{\row,\col} < \frac{2}{3} + 2z + \frac{1}{3} + z = 1 + 3z$. 
\end{proof}

We now use this proposition, along with Markov's inequality, to prove the bound
for $O_\row$ specified in Proposition~\ref{prop:inequalities}.

\begin{proposition}
\label{prop:point1}
If $\row$ is a $q$-bad row, then
$\sum_{\col \in O_{\row}} \y_{\col} \le \frac{2qz}{\frac{1}{3} - 2z}$.
\end{proposition}
\begin{proof}
Consider the random variable $T = \frac{4}{3} + z - R_{\row,\col} - C_{\row,\col}$,
where $\row$ is fixed and $\col$ is sampled from $\y$.
From Equation~\eqref{eqn:sum}, we have that $T$ takes values in the range
$[0, \frac{4}{3} + z]$. 
Utilizing Proposition~\ref{prop:bad2}, part 1, along with
Equation~\eqref{eqn:cavg} gives the following:
\begin{equation*}
R_{\row}\y + C_{\row}\y \geq \frac{4}{3} + (1-2q)z.
\end{equation*}
Therefore, we have the following expression for the expectation of $T$:
\begin{align*}
E[T] &   = \frac{4}{3} + z - E_{\col\sim\y}[R_{\row,\col} + C_{\row,\col}] \\
     & \le \frac{4}{3} + z - \left(\frac{4}{3} + (1-2q)z\right) = 2qz
\end{align*}
By Proposition~\ref{prop:farsum}, for each $\col \in O_{\row}$, we have
$R_{\row,\col} + C_{\row,\col} \le 1 + 3z$. Hence, we have $T \ge \frac{4}{3} + z -
(1 + 3z) = \frac{1}{3} - 2z$ for each $\col \in O_{\row}$. Therefore, we must have
$\prob(T \ge \frac{1}{3} - 2z) \geq \sum_{\col \in O_{\row}} \y_{\col}$.
Applying Markov's inequality completes the proof:
\begin{equation*}
\prob(T \ge \frac{1}{3} - 2z) \le \frac{E[T]}{\frac{1}{3} - 2z} \le \frac{2qz}{\frac{1}{3} - 2z}.
\end{equation*}
\end{proof}

Now we prove the inequality that was given for $B_\row$ in
Proposition~\ref{prop:inequalities}.

\begin{proposition}
\label{prop:blower}
If $\row$ is a $q$-bad row, then $\sum_{\col \in B_{\row}} \y_{\col} \ge \frac{\frac{1}{3} + z -
qz - (\frac{1}{3} + z) \sum_{\col \in O_{\row}} \y_{\col}}{\frac{2}{3} - z}.$
\end{proposition}

\begin{proof}
Since the sets $B_{\row}$, $S_{\row}$, and $O_{\row}$ are disjoint, we can
write Definition~\ref{def:ravg} as:
\begin{equation*}
\sum_{\col \in B_{\row}} \y_{\col} R_{\row,\col} + \sum_{\col \in S_{\row}} \y_{\col} R_{\row,\col}
+ \sum_{\col \in O_{\row}} \y_{\col} R_{\row,\col} \geq \frac{2}{3}+2z-qz.
\end{equation*}
We know that $R_{\row,\col}\le 1$ for each $\col\in B_{\row}$, that $R_{\row,\col} \le
\frac{2}{3} + 2z$ for each $\col\in O_{\row}$, and that $R_{\row,\col} \le \frac{1}{3} + z$
for each $\col\in S_{\row}$. Therefore we obtain the following inequality:
\begin{equation*}
1 \cdot \sum_{\col\in B_{\row}} \y_{\col} + \left(\frac{1}{3} + z\right) \cdot \sum_{\col\in S_{\row}} \y_{\col}
+ \left(\frac{2}{3} + 2z\right) \cdot \sum_{\col\in O_{\row}} \y_{\col} \ge \frac{2}{3} + 2z - qz.
\end{equation*}
Furthermore, since
$\sum_{\col\in S_{\row}} \y_{\col} = 1 - \sum_{\col\in B_{\row}} \y_{\col}
 - \sum_{\col\in O_{\row}} \y_{\col}$, we have:
\begin{equation*}
\sum_{\col\in B_{\row}} \y_{\col}
 + \left(\frac{1}{3}+z\right) \cdot \left(1 - \sum_{\col\in B_{\row}} \y_{\col} - \sum_{\col\in O_{\row}} \y_{\col}\right)
 + \left(\frac{2}{3} + 2z\right) \sum_{\col\in O_{\row}} \y_{\col} \ge \frac{2}{3} + 2z - qz.
\end{equation*}
Rearranging this gives:
\begin{equation*}
\left(\frac{2}{3} - z\right) \cdot \sum_{\col\in B_{\row}} \y_{\col} \ge \frac{1}{3} + z -qz -
\left(\frac{1}{3} + z\right) \sum_{\col\in O_{\row}} \y_{\col}.
\end{equation*}
Finally, this allows us to conclude that:
\begin{equation*}
\sum_{\col\in B_{\row}} \y_{\col} \ge \frac{\frac{1}{3} + z - qz -
(\frac{1}{3} + z) \sum_{\col\in O_{\row}} \y_{\col}}{\frac{2}{3} - z}.  
\end{equation*}
\end{proof}

Finally, we prove the inequality that was given for $S_j$ in
Proposition~\ref{prop:inequalities}. This proof is very similar to the proof of
Proposition~\ref{prop:blower}, except that we substitute into
Equation~\eqref{eqn:cavg} rather than Definition~\ref{def:ravg}.

\begin{proposition}
\label{prop:slower}
If $\row$ is a $q$-bad row then $\sum_{\col\in S_{\row}} \y_{\col} \ge \frac{\frac{1}{3} -2z-qz
- (\frac{1}{3} + z)\sum_{\col\in O_{\row}} \y_{\col}}{\frac{2}{3} - z}.$
\end{proposition}
\begin{proof}
Since the sets $B_{\row}$, $S_{\row}$, and $O_{\row}$ are disjoint, we can rewrite
Equation~\eqref{eqn:cavg} as:
\begin{equation*}
\sum_{\col \in B_{\row}} \y_{\col} C_{\row,\col} + \sum_{\col \in S_{\row}} \y_{\col} C_{\row,\col}
+ \sum_{\col \in O_{\row}} \y_{\col} C_{\row,\col} \geq \frac{2}{3}-z-qz.
\end{equation*}
We know that $C_{\row,\col}\le 1$ for each $\col\in S_{\row}$, that $C_{\row,\col} \le
\frac{2}{3} + 2z$ for each $\col\in O_{\row}$, and that $C_{\row,\col} \le \frac{1}{3} + z$
for each $\col\in B_{\row}$. Therefore we obtain the following inequality:
\begin{equation*}
1 \cdot \sum_{\col\in S_{\row}} \y_{\col} + \left(\frac{1}{3} + z\right) \cdot \sum_{\col\in B_{\row}} \y_{\col}
+ \left(\frac{2}{3} + 2z\right) \cdot \sum_{\col\in O_{\row}} \y_{\col} \ge \frac{2}{3} - z - qz.
\end{equation*}
Furthermore, since 
$\sum_{\col\in B_{\row}} \y_{\col} = 1 - \sum_{\col\in S_{\row}} \y_{\col}
 - \sum_{\col\in O_{\row}} \y_{\col}$, we have:
\begin{equation*}
\sum_{\col\in S_{\row}} \y_{\col}
 + \left(\frac{1}{3}+z\right) \cdot \left(1 - \sum_{\col\in S_{\row}} \y_{\col} - \sum_{\col\in O_{\row}} \y_{\col}\right)
 + \left(\frac{2}{3} + 2z\right) \sum_{\col\in O_{\row}} \y_{\col} \ge \frac{2}{3} - z - qz.
\end{equation*}
Rearranging this gives:
\begin{equation*}
\left(\frac{2}{3} - z\right) \cdot \sum_{\col\in S_{\row}} \y_{\col} \ge \frac{1}{3} -2z -qz -
\left(\frac{1}{3} + z\right) \sum_{\col\in O_{\row}} \y_{\col}.
\end{equation*}
Finally, this allows us to conclude that:
\begin{equation*}
\sum_{\col\in S_{\row}} \y_{\col} \ge \frac{\frac{1}{3} - 2z - qz -
(\frac{1}{3} + z) \sum_{\col\in O_{\row}} \y_{\col}}{\frac{2}{3} - z}.  
\end{equation*}
\end{proof}

Now that we have shown Propositions~\ref{prop:point1},~\ref{prop:blower},
and~\ref{prop:slower}, we have completed the proof of
Proposition~\ref{prop:inequalities}

\newpage
\section{Mathematica Code}

\begin{lstlisting}[language=Mathematica]
(* The function phi(z, qbar) from the paper *)
Phi[z_,qbar_] := 1 + (1/3+z+qbar*z)/(1/3 - 2*z - qbar*z - (1/3+z)*(2*qbar*z)/(1/3-2*z))
	
(* Solves the LP from the paper for the given set of parameters *)
LP[z_, t_, version_] := Maximize[{

	(* ========================  Objective  =============================*)
	(1 - t) * (2/3 + 2*z - q*z) 
	+ t*(
		Phi[z, 3]
		*
		(sb + (1/3 + z) * ss + (2/3 + 2*z)*so)
		+
		 ob + (1/3 + z) * os + (2/3 + 2*z)*oo
	),

	(* ========================  Constraints  ===========================*)

	(* Constraints from the paper *)
	bb + bs + bo >= (1/3 + z - qbar*z - (1/3 + z)*(ob + os + oo))/(2/3 - z),
	bb + sb + ob >= (1/3 + z - q*z - (1/3 + z)*(bo + so + oo))/(2/3 - z),
	sb + ss + so >= (1/3 - 2*z - qbar*z - (1/3 + z)*(ob + os + oo))/(2/3 - z),
	bs + ss + os >= (1/3 - 2*z - q*z - (1/3 + z)*(bo + so + oo))/(2/3 - z),
	ob + os + oo <= 2*qbar*z/(1/3 - 2*z),
	bo + so + oo <= 2*q*z/(1/3 - 2*z),

	(* Either bs = 0 or sb = 0 *)
	version == 0,
	
	(* Sum to one *)
	bb + bs + bo + sb + ss + so + ob + os + oo == 1,

	(* qbar constraints *)
	qbar <= 3,
	qbar <= q,

	(* Non-negativities *)
	bb >= 0,
	bs >= 0,
	bo >= 0,
	sb >= 0,
	ss >= 0,
	so >= 0,
	ob >= 0,
	os >= 0,
	oo >= 0,
	q >= 0,
	qbar >= 0

	},

	(* ========================  Variables  ============================ *)
	{bb, bs, bo, sb, ss, so, ob, os, oo, q, qbar}
]

(* For a given z and k, this function uses grid search to find the t that
 minimizes the objective function of LP(z, t, k) *)
FindT[z_, version_] := Module[{t, tstep, best, tbest, outcome},
	t = 0;
	tstep = 1/1000;
	best = 1;
	tbest = 0;
	While[t <= 2/10,
		outcome = LP[z, t, version][[1]];
		If[outcome < best, best = outcome; tbest = t];
		t += tstep
	];
	{best, tbest}
]

(* This function uses grid search to find the best witness (z, t_0, t_1). We
   have chosen starting parameters that give the result from the paper in a
   reasonable amount of time *)
Module[{z, zstep, zbest, tbest, outcome},
	(* Starting z. This can be reduced to 0 if desired, but the program will
	 take a very long time if zstep is not increased. *)
	z = 59137/10000000;
	zstep = 1/1000000000;
	zbest = 0;
	t1best = -1;
	t2best = -1;
	(* The upper bound on z can be increased to any number strictly less than
	 (13-3*Sqrt[17])/24, but the program will take a very long time if zstep is
	 not increased *)
	While[z <= 59138/10000000,
		z = z + zstep;
		out1 = FindT[z, sb];
		out2 = FindT[z, bs];
		outcome = Max[out1[[1]], out2[[1]]];
		If[outcome <= 2/3-z, zbest = z; t1best = out1[[2]]; t2best = out2[[2]]]
	];
	(* Print out the exact rational z *)
	Print[zbest]
	(* Print out the witness in decimal form *)
	Print[N[zbest]];
	Print[N[t2best]];
	Print[N[t1best]];
]
\end{lstlisting}

\end{document}